%Paper: astro-ph/9407034
%From: karsten@ulysses.llnl.gov (Karsten Jedamzik)
%Date: Tue, 12 Jul 94 12:57:49 PDT

\lineskip=3pt minus 2pt
\lineskiplimit=3pt
\magnification=1200
\centerline{\bf ON CONSTRAINING ELECTROWEAK-BARYOGENESIS}
\vskip 0.08in
\centerline{\bf WITH INHOMOGENOUS PRIMORDIAL NUCLEOSYNTHESIS}
\vskip 0.4in
\baselineskip=24pt plus 2pt
\centerline{\sl G. M. Fuller$^1$, K. Jedamzik$^1$, G. J. Mathews$^2$}
\vskip 0.1in
\centerline{and}
\vskip 0.1in
\centerline{\sl A. Olinto$^3$}
\vskip 0.75in
\centerline{\bf ABSTRACT}
\vskip 0.06in
Primordial nucleosynthesis
calculations are shown to be able to provide constraints on models of
electroweak baryogenesis which produce a highly inhomogeneous distribution
of the baryon-to-photon ratio.
Such baryogenesis scenarios
overproduce $^4$He and/or $^7$Li
and can be ruled out whenever
a fraction $f ^<_\sim 3\times 10^{-6}$(100 GeV/$T)^3$
of nucleated bubbles of broken-symmetry phase
contributes $^>_\sim$10\% of the baryon number within a horizon volume.

\vskip 1.in
\baselineskip=14pt plus 2pt
\noindent
$^1$ Department of Physics, University of California, San Diego,
La Jolla CA 92037-0319

\noindent
$^2$ Physics Department, University of California, Lawrence Livermore
National Laboratory, Livermore, CA 94550

\noindent
$^3$ Astronomy and Astrophysics, The University of Chicago, Chicago,
IL 60637

\vfil\eject

\baselineskip=24pt plus 2pt

In this letter we discuss how the sensitivity of big bang
nucleosynthesis to the baryon-to-photon number ($\eta$)
and its spatial distribution could be utilized to probe electroweak
physics in a new manner. Models of electroweak baryogenesis have
so far concentrated on producing the presently observed value for $\eta$.  In
this paper, we point out a new constraint on models that
result in a highly inhomogeneous distribution of $\eta$.
We will show that
micro-physical processes that generate fluctuations in $\eta$ on
sub-horizon scales for epochs corresponding to temperatures $T$
$^<_{\sim}$ 1.5 TeV may be subject to nucleosynthesis constraints.

A long-standing problem in astrophysics is the explanation for the
apparent baryon number asymmetry in the universe. Reference [1]
provides an overview of this problem and the attempts to solve it.
Ever since the work of Sakharov [2] an explanation has been sought
for the baryon number asymmetry in C, CP, and baryon-number violating processes
in environments
associated with departures from thermal and chemical equilibrium in
the early universe. However,
any net baryon number generated at very early epochs in the history
of the universe (e.g., via C and CP violating, nonequilibrium
baryon number violating decay
of heavy $X$ and $Y$ bosons associated with Grand
Unification) will probably, though not necessarily inevitably,
be erased by subsequent anomalous electroweak processes [3].
Regeneration of baryon number could then occur during a first order cosmic
electroweak
symmetry-breaking phase transition [3].
It is not clear if adequate baryogenesis could be achieved
with a minimal Weinberg-Salam model without implying a
Higgs mass below the present experimental lower
bound (see, however, reference [4]). Several plausible
extensions of the minimal standard model,
such as multi-Higgs models or supersymmetric models,
could lead to
significant baryon number generation at this epoch [5,6,7].
A review of baryogenesis associated with first-order
electroweak phase transitions is given in reference [8].

Since a temperature dependent nucleation rate is a
generic feature of first order phase transitions, we expect some
supercooling in a primordial electroweak transition and the
concomitant generation of distinct bubbles of low temperature phase.
These bubbles of broken phase grow until they coalesce. As the
bubble walls propagate toward coalescence the universe is out of
thermal and chemical equilibrium in the vicinity of the walls.
These nonequilibrium conditions, together with baryon
number violating anomalous electroweak interactions and C and
CP-violation, provide all the necessary ingredients for baryogenesis.
The necessity of this baryogenesis occurring in the
inhomogeneous environments engendered by bubble nucleation and
coalescence ultimately may lead to an inhomogeneous distribution of
$\eta$ and, hence, entropy-per-baryon.

In most of these baryogenesis scenarios the final distribution in
$\eta$ is probably too homogeneous to affect nucleosynthesis. However, one can
speculate on models in which significant inhomogeneities in $\eta$,
($\Delta\eta /\eta {{}_{\sim}^>} 1$), may occur. Such inhomogeneities, for
example,
might arise in nonadiabatic (thin wall) models [9] whenever the velocity of an
expanding broken phase bubble varies during the transition. In nonadiabatic
scenarios the rate of baryogenesis depends strongly upon the velocity of the
wall. We note that this velocity may change at only about the 10\% level during
the course of the transition. However, there remains considerable uncertainty
in the determination of the wall velocity in these models. As recently pointed
out, this effect may occur in adiabatic models as well [10].

Another possibility might be the formation [11] of distinct domains of
baryon-number and anti-baryon number. After annihilating they could leave
behind a small number of baryon bubbles containing all of the net baryon
number. Finally, strong spatial inhomogeneities may result from any scenario in
which most of the generated net baryon number is associated with the collisions
of bubble walls at the end of the transition. Although such models are
speculative it is nevertheless interesting to investigate the constraints which
might be placed on such scenarios from primordial nucleosynthesis.

If fluctuations occur, the bubble size at coalescence will probably provide a
typical
length scale of fluctuations in $\eta$, but fluctuations
 can occur on larger scales than that. However, it is difficult to quantify
that length scale. Thermal and/or quantum nucleation is especially
difficult to follow at the electroweak epoch because the nucleating
action may be dynamically renormalized by the presence of bubbles of
broken phase [12]. Another complication may be hydrodynamic instability of
phase boundaries[13].

Despite these caveats it is nevertheless instructive to consider
simple models of homogeneous nucleation of phase in the small
supercooling limit [14]. In these models the nucleation rate per unit
volume is assumed to be,
$$p(T)\approx CT^4e^{-S(T)}\ ,\eqno(1)$$
where $S(T)=a(T)\bigl(T_c/(T_c-T)\bigr)$ is the nucleating action,
with  $a(T)$ a monotonically
increasing function of temperature, and
where $C$ is a scale factor of order unity.
Integrating the nucleation rate
through the end of the phase transition, and assuming that
bubble walls move at the speed of light, yields an estimate
for the time required for bubbles to coalesce. We can express this
coalescence time (or bubble size at coalescence) as a fraction $\delta$ of the
Hubble time
(or horizon scale) $H^{-1}$ [13],
$$\delta\approx\biggl(4B{\rm ln}\Bigl({m_{pl}\over
T_c}\Bigr)\biggr)^{-1}\ ,\eqno(2)$$
where $B$ is the logarithmic derivative of the nucleating action
$S$, in units of $H^{-1}$ at the epoch of the phase transition. The
value of $B$ depends on calculable details of models for the
electroweak transition and is within one or two orders of magnitude
of unity. The horizon size is,
$$H^{-1}\approx \Bigl({90\over
8\pi^3}\Bigr)^{1/2}g^{-1/2}{m_{pl}\over T^2}\approx (1.45
cm)\Bigl({g\over 100}\Bigr)^{-1/2}\Bigl({T\over 100{\rm
GeV}}\Bigr)^{-2}\ ,\eqno(3)$$
where $g=\sum_bg_b+7/8\sum_fg_f$ is the total statistical
weight in relativistic bosons $(g_b)$ and fermions $(g_f)$ at
temperature $T$, and $m_{pl}$ is the Planck mass. In the standard
model, $g\approx 100$ for an electroweak transition at $T=100$ GeV.
The total statistical weight is slightly uncertain due to the
unknown top quark mass and extra degrees of freedom associated with
extensions of the standard model.

The average bubble size at coalescence $\delta H^{-1}$ is a result of
competition between the
nucleation rate and the very slow expansion of the universe. Most
bubbles will have size $\delta H^{-1}$ at coalescence. This follows
on noting that larger bubbles would have to be nucleated early, near
$T_c$, where the nucleation rate is exponentially suppressed. Smaller
bubbles would have to be nucleated near the end of the phase
transition where the effective nucleation rate is again small, since
very little unbroken phase would remain.
A nucleation/coalescence epoch which approximates the homogeneous
nucleation scenario will leave a nearly regular lattice of bubbles at
coalescence.

Even though several electroweak scenarios have been proposed [8,9,15]
not much is known about the actual nucleation scale $\delta H^{-1}$ and the
expected coexistence temperature $T_c$ in these models. It has been
argued that the minimal standard model gives $\delta\sim 10^{-3}$
[16].
These models have not been
investigated in sufficient detail to ascertain the relationship between
the nucleation scale $\delta H^{-1}$ and the scale of separation between
centers of fluctuations
in $\eta$, which we shall denote $\delta_{\eta} H^{-1}$.
However, it is possible that
$\delta_{\eta}$ $^>_{\sim}$ $\delta$, corresponding to less than, or equal to,
one fluctuation produced per nucleated bubble.

In Ref.[17], two of us (hereafter referred as JF) have studied in detail the
evolution of fluctuations from $T \approx 100$ MeV to $T \approx 1$ keV taking
into account
neutrino, photon, and baryon dissipation processes. JF's results
indicate that fluctuations generated at the electroweak epoch may survive
through the epoch of primordial nucleosynthesis. If fluctuations with
particular
characteristics produced at an early epoch did survive, their presence could
alter the nuclear abundance yields emerging from primordial nucleosynthesis
[18,19]. If these abundance yields do not agree with observationally inferred
primordial abundances, then we can conclude that these fluctuations could not
have existed. This, in turn, would allow us to constrain the fluctuation
generation mechanism.

We follow JF and define the amplitude of fluctuations $\Delta (x)$, in terms
of the spatial distribution of baryon-to-photon number, $\eta (x)$, and its
horizon average, $\bar\eta$, by $\eta (x)={\bar\eta}\bigl(1+\Delta (x)\bigr)$.
The corresponding distribution in entropy-per-baryon is then
$s(x)={\bar s}\bigl(1+\Delta (x)\bigr)^{-1}$, where the average conserved
entropy-per-baryon in units of Boltzmann's constant is
${\bar s}\approx 2.63\times 10^8\Omega_b^{-1} h^{-2}$.
In this expression $\Omega_b$ is the fraction of the closure density
contributed
by baryons and $h$ is the present Hubble parameter in units of 100 km s$^{-1}$
Mpc$^{-1}$.  In this paper $\Delta(x)$ always refers to the {\it initial}
amplitude
of the fluctuations.

The primary criterion for fluctuation survival is that the scale
associated with the separations of the centers of fluctuations
($\delta_{\eta}H^{-1}$) be comparable to, or exceed,
the comoving proton
diffusion length ($d_{100}$) at the beginning of the nucleosynthesis epoch
[17,18,19]. Here $d_{100}$ is the comoving proton diffusion length referenced
to the epoch of $T=100$ GeV (see for example eq.4).
Were this condition not satisfied, baryon diffusion would erase fluctuations in
$\eta$ prior to nucleosynthesis.
The proton diffusion length is actually a fairly sensitive function of
amplitude
$(1+\Delta )$. In Figure 1 we give the comoving proton diffusion length
$d_{100}$ at the
epoch $T=500$ keV as a function of $(1+\Delta )$.
This temperature very roughly corresponds to the epoch of weak
freeze-out, where the neutron-to-proton interconversion rate from lepton
capture
falls below the free neutron decay rate.
Note that higher baryon density
implies a smaller diffusion length for baryons. Whenever $(1 + \Delta)
^<_\sim 10^2$
the baryon diffusion length corresponds to
$d_{100}\sim 0.1$ cm.

We also can describe fluctuations by their separation length scale, $l_{100}$,
where we express a length scale co-moving with the Hubble expansion in terms of
its proper length at an epoch where $T=100$ GeV. The corresponding proper
length at any epoch where the temperature is $T$ is then
$$l=l_{100}\biggl({R\over
R_{100}}\biggr)=l_{100}\biggl({g_{100}^{1/3}T_{100}
\over g^{1/3}T}\biggr)\ ,
\eqno(4)$$
where $R$ and $R_{100}$ are the scale factors at an epoch of temperature $T$
and
100 GeV, respectively, $T_{100}=100$ GeV, and where $g$ and $g_{100}$ are the
statistical weights in relativistic particles at an epoch of temperature $T$
and
100 GeV, respectively. In this expression we have assumed that the
co-moving entropy density is conserved.

In order for a fluctuation to affect the outcome of nucleosynthesis $l_{100}$
$^>_{\sim}$  $l_{100}^{min} \approx d_{100}$. This scale is found from detailed
nucleosynthesis calculations to be roughly the scale of the proton diffusion
length at the nucleosynthesis epoch. Physically, the origin of this limiting
length is that any fluctuation scale smaller than the proton diffusion length
will be damped out by baryon diffusion prior to nucleosynthesis.
Therefore, the minimum fluctuation scale for inhomogeneous nucleosynthesis
effects can be expressed in terms of a fraction of the horizon scale $H^{-1}$
at
any epoch as $$\delta_{min}\equiv {l^{min}\over H^{-1}}\approx
l_{100}^{min}\biggl({8\pi^3\over
90}\biggr)^{1/2}g_{100}^{1/3}g^{1/6}{T
T_{100}\over m_{pl}}\ ,\eqno(5a)$$
$$\delta_{min}\approx (6.9\times 10^{-2})\Bigl({g_{100}\over
100}\Bigr)^{1/2}\Bigl({g\over g_{100}}\Bigr)^{1/6}\Bigl({l_{100}^{min}\over 0.1
{\rm cm}}\Bigr)\Bigl({T\over 100 {\rm GeV}}\Bigr).\eqno(5b)$$
Note that $\delta_{min}$ $^<_{\sim}$ 1 for $T$ $^<_{\sim}$ 1.45 TeV.
We conclude that micro-physical, subhorizon-scale fluctuation-generating
processes operating at epochs for which $T$ $^<_{\sim}$ 1.45 TeV
conceivably could have
constrainable nucleosynthesis signatures.
Fluctuations in $\eta$ on initially super-horizon scales $l$ which satisfy $l$
$^>_{\sim}$ $l_{min}$ are similarly at risk of running afoul of primordial
abundance constraints. We note, however, that even fluctuations with $l$
$^<_{\sim} l_{min}$ may yet survive to affect nucleosynthesis if they have
amplitudes large enough that the length scales of their high density regions,
$l_{100}^H$, exceed
$d_{100}$. In this case baryons would be unable to diffuse out of the high
density cores of fluctuations prior to nucleosynthesis.

In any scheme for baryogenesis associated with an electroweak symmetry breaking
epoch at temperature $T$ we must produce the average proper baryon
number density within the horizon
$${\bar n}_b={S\over {\bar s}}\approx 0.167 {\rm GeV^3}\biggl({g\over
100}\biggr
)
\biggl({T\over 100 {\rm GeV}}\biggr)^3\Omega_b h^2,\eqno(6)$$
where the entropy per unit proper volume is
$S\approx (2\pi^2/45)gT^3$. Homogeneous and inhomogeneous standard big bang
nucleosynthesis calculations together with observational abundance constraints
imply that $\Omega_b\approx 0.01 h^{-2}$ [1,18].

Assume that baryons are distributed in high density regions with baryon number
density
$n_b^H$, which in total occupy a fraction $f_V$ of the horizon volume, and in
low density regions with baryon number density $n_b^L$. In this case, we can
write
$${\bar n}_b=f_Vn_b^H+(1-f_V)n_b^L.\eqno(7)$$
We define $\Lambda_H\equiv n_b^H/{\bar n}_b$ and $\Lambda_L\equiv n_b^L/{\bar
n}_b$, so that the density contrast between high and low density regions is
$\Lambda\equiv \Lambda_H/\Lambda_L$. If the horizon is filled with a regular
lattice of fluctuation cells whose centers are separated by $l_{100}^s$, then
the length scale of high density regions is $l_{100}^H=f_V^{1/3}l_{100}^s$.

\vskip 0.15in
In Ref.[18], three of us have calculated in detail the outcome of primordial
nucleosynthesis with inhomogeneous initial conditions. In these calculations
the nuclear reaction rates were coupled to all significant fluctuation
dissipation processes: neutrino heat transport, baryon diffusion, photon
diffusive heat transport, and hydrodynamic expansion with photon-electron
Thomson drag. The light element abundance yields are found to be inconsistent
with observations for all but a very narrow range of fluctuation
characteristics. This is why nucleosynthesis is so powerful in constraining
primordial inhomogeneities.

A representative case of  these
calculations  is displayed in Figure 2. In this figure we show the $^4$He mass
fraction and number fraction of $^7$Li emerging from an inhomogeneous big bang
with  fluctuation separations $l_{100}^s$. Dotted lines indicate
$l_{100}^{min}$
(BDL) and the electroweak horizon scale (EWH). Though we show results for
particular values of fluctuation amplitude and gaussian width $a_{100}$
(roughly,
$f_V\approx (a_{100}/l_{100}^s)^3)$, the figure illustrates some general trends
for  abundance yields as a function of length scale. In particular, we note
that
when $l_{100}^{min}$ $^<_{\sim}$ $l_{100}^s$ $^<_{\sim}$ EWH the abundances of
$^4$He and/or $^7$Li always exceed observational limits [20]. This is a general
feature of inhomogeneous primordial nucleosynthesis whenever $l_{100}^s$ is
below the electroweak horizon scale.  It is, however, intriguing that the
electroweak horizon is close to the minimum in $^4$He and $^7$Li (the \lq\lq
helium dip\rq\rq).

We have also explored inhomogeneous nucleosynthesis yields for the light
elements as functions of initial density contrast
and volume filling fraction, $\Lambda$ and $f_V$, respectively.
Figure 3
shows the results of numerous numerical calculations for $l_{100}^s=0.5$ cm. In
this figure
the parameter space of $\Lambda$ and $f_V$ laying to the right of the shaded
line gives
$^4$He overproduction ($^4$He mass fraction $> 24$\%).
In general, we find that $^4$He
and/or $^7$Li are overproduced whenever 10\% or more of the baryons reside in
high density regions and either $l_{100}^s$ $^>_{\sim}$ $l_{100}^{min}$ or
$l_{100}^H$ $^>_{\sim}$ $d_{100}$.

This constraint can be put in the context of electroweak baryogenesis with a
simple model. Assume that an electroweak phase transition has produced a
regular
lattice of bubbles at coalescence, all of equal size. In fact, we expect
a distribution of bubble sizes, but for now assume all bubbles
have size equal to the nucleation
scale $\delta H^{-1}$. Since we expect $\delta\sim 10^{-3}$ there will be
roughly
$\delta^{-3}\sim 10^9$ bubbles within the horizon.
As they expand toward coalescence, these $10^9$ bubbles must produce
the horizon-averaged baryon number density ${\bar n}_b$.
It is likely, however, as noted above that not
all bubbles will contribute equally to this average. Assume that a fraction
$f$ of
the bubbles contributes a substantial fraction $b$ of the total baryon number
in
the horizon. As far as the net baryon distribution is concerned, this scenario
approximates a regular lattice of fluctuations with effective horizon-fraction
separations $\delta_{\eta}\approx f^{-1/3}\delta$.

Equation (7) shows that the baryon distribution is characterized by two
independent quantities, which we take to be $f_V$ and $\Lambda$.
In the above
hypothetical scenario it is clear that $f_V\leq f$, with equality obtaining
when the baryon number produced in a bubble is uniformly distributed over the
volume swept out by the bubble wall. The density contrast will be
$$\Lambda\geq\biggl({{1-f_V}\over f_V}\biggr)\biggl({b\over {1-b}}\biggr)\
,\eqno(8)$$
assuming that each of the $f\delta^{-3}$ bubbles contributes equally to $b{\bar
n}_b$. Equality in equation (8) obtains in the limit of uniform baryon
distribution across each of the \lq\lq significant\rq\rq $f\delta^{-3}$
bubbles.
Note that we also assume that significant bubbles are uniformly
distributed in space.
In this model, $b=(f_Vn_b^H)/{\bar n}_b=f_V\Lambda_H$.
Since $\Lambda_H=1+\Delta$, we may
conclude that $(1+\Delta)\approx b/f_V$.
The total fraction of baryons in high density regions is $f_V\Lambda$, which is
just $b$ in the limit where $\Lambda f_V<< 1$. Our nucleosynthesis constraint
will apply whenever more than 10\% of the baryons are in high density regions,
$b$ $^>_{\sim}$ $0.1$,  and either the fluctuation separation exceeds the
minimum value
required for nucleosynthesis effects,
$f^{-1/3}\delta >\delta_{min}$,
or  the high density region length scale exceeds the proton diffusion
length,$\ l_{100}^H$ $^>_{\sim}$ $d_{100}$.

For example, in the context of this model assume that only 1000 bubbles out of
the total of $10^9$ bubbles in the horizon produce 90\%
of the baryon number. This implies that $f_V\approx f\approx 10^{-6}$ and
$\delta_{\eta}\approx f_V^{-1/3}\delta\approx 10^2\delta \approx 0.1 >
\delta_{min}$. This last comparison follows on assuming that $\delta \approx
10^{-3}$.
Equation (8) implies that $\Lambda\approx 9\times 10^6$. Since
$b=0.9$ exceeds 0.1 and $\delta_{\eta}>\delta_{min}$ we would conclude that
this
scenario is incompatible with nucleosynthesis constraints and, therefore,
ruled out.
In general, whenever $f<(\delta/\delta_{min})^3\sim 3\times
10^{-6}(100{\rm GeV}/T)^3$ and $b>0.1$ the nucleosynthesis
constraint will be violated.  In other words, any scenarios where
fewer than about 3000 bubbles
contribute more than 10\% of the baryon number can be ruled out.

In some models of electroweak baryogenesis the significant bubbles
for baryon production might be those which are nucleated earliest.
We would then expect these bubbles to be larger at coalescence than
the average bubble ($\delta H^{-1}$). If the ratio of significant
bubble size to the average bubble size is $r$, and there are
$f\delta^{-3}$ significant bubbles, then the effective volume
filling fraction for the higher density regions of the baryon
distribution is roughly $f_V$ $^<_{\sim}$ $r^3f$. Note that in this
case, however, the effective horizon-fraction separation is still
$\delta_{\eta}\approx f^{-1/3}\delta$.

Though the detailed relationship between the fluctuation scale and amplitude
and
such model parameters as the Higgs particle mass, the top quark mass, and the
critical temperature $T_c$ are as yet poorly understood, they are in principle
calculable. The parameters of electroweak baryogenesis models must not lead
to a violation of the constraints on
fluctuation characteristics as describes above.
This is probably readily attainable
for some models, but may represent a restriction for others.

\vskip 0.15in
\centerline{\bf Acknowledgements}
We wish to thank N. J. Snyderman and H. Kurki-Suonio and M.S. Turner for many
useful
conversations. We also wish to thank an anonymous referee for many useful
comments regarding a preliminary version of this manuscript. This work was
supported in part by NSF Grant
PHY91-21623, IGPP LLNL Grant 93-22, and a California Space Institute
Grant. It was also performed in part under the auspices of the U.S.
Department of Energy by the Lawrence Livermore National Laboratory
under contract number W-7405-ENG-48 and DoE Nuclear Theory Grant
SF-ENG-48 and by the University of Chicago under DOE grant FG02-91ER40606.

\vfill\eject

\centerline{\bf References}

\noindent
{\bf [1]} E. W. Kolb and M. S. Turner, \lq\lq\ The Early Universe \rq\rq\ ,
Addison-Wesley (1990).

\noindent
{\bf [2]} A. D. Sakharov, JETP Lett. 5 (1967) 24.

\noindent
{\bf [3]} V. A. Kuzmin, V. A. Rubakov, and M. E. Shaposhnikov, Phys. Lett.
155B (1985) 36; E. W. Kolb and M. S. Turner Mod. Phys. Lett. A2
(1987) 285.

\noindent
{\bf [4]} M. E. Shaposhnikov, Phys. Lett. B277 (1992) 324.

\noindent
{\bf [5]} M. Dine, P. Huet, R. Singleton, and L. Susskind, Phys. Lett.
257B (1991) 351; M. Dine, O. Lechtenfeld, B. Sakita, W. Fischler,
and J. Polchinski, Nucl. Phys. B342 (1990) 381.

\noindent
{\bf [6]} A. G. Cohen, D. B. Kaplan, and A. E. Nelson, Phys. Lett. 245B
(1990) 561; A. G. Cohen, D. B. Kaplan, and A. E. Nelson, Phys. Lett.
B263 (1991) 86;
A. G. Cohen, D. B. Kaplan, and A. E. Nelson, Nucl. Phys.
B349 (1991) 727.

\noindent
{\bf [7]} L. McLerran, M. Shaposhnikov, N. Turok, and M. Voloshin,
Phys. Lett. B256 (1991) 451; N. Turok and J. Zadrozny, Nucl. Phys.
B369 (1992) 729.

\noindent
{\bf [8]} A. G. Cohen, D. B. Kaplan, and A. E. Nelson, Ann. Rev.
Nucl. and Part. Sci. in press.

\noindent
{\bf [9]} A. E. Nelson, D. B. Kaplan, and A. G. Cohen, Nucl. Phys.
B373 (1992) 453.

\noindent
{\bf [10]} M. Joyce, T. Prokopec, and N. Turok, preprint hep-ph/9401352.

\noindent
{\bf [11]} D. Comelli and M. Pietroni, Phys. Lett. B 306 (1993) 67;
J.R. Espinosa and M. Quiros, Phys. Lett. B 307 (1993) 106; D. Comelli,
M. Pietroni, and A. Riotto, Nucl. Phys. B 412 (1994) 441.

\noindent
{\bf [12]} M. Gleiser and E. W. Kolb, Phys. Rev. Lett. 69 (1992)
1304.

\noindent
{\bf [13]} K. Freese and M. Kamionkowski, Phys. Rev. Lett. 69 (1992)
2743; P. Huet, K. Kajantie, R. Leigh, B. Liu, and L. McLerran,
preprint SLAC-PUB-5943.

\noindent
{\bf [14]} C. Hogan, Phys. Lett. 133B (1983) 172; S. Coleman, Phys. Rev.
D15 (1977) 2929.

\noindent
{\bf [15]} M. Dine, R. G. Leigh, P. Huet, A. Linde, and D. Linde,
Phys. Rev. D46 (1992) 550; K. Enqvist, J. Ignatius, K. Kajantie, and
K. Rummukainen, Phys. Rev. D45 (1992) 3415; B. H. Liu, L. McLerran,
and N. Turok, Phys. Rev. D46 (1992) 2668.

\noindent
{\bf [16]} G. W. Anderson and L. J. Hall, Phys. Rev. D45 (1992) 2685.

\noindent
{\bf [17]} K. Jedamzik and G. M. Fuller, Astrophys. J. 423 (1994) 33.

\noindent
{\bf [18]} K. Jedamzik, G. M. Fuller, and G. J. Mathews, Astrophys.
J. 423 (1994) 50.

\noindent
{\bf [19]} J. H. Applegate, C. J. Hogan, and R. J. Scherrer, Phys. Rev.
D35 (1987) 1151; C. R. Alcock, G. M. Fuller, and G. J. Mathews,
Astrophys. J. 320 (1987) 439; See the review by R. A. Malaney and G.
J. Mathews, Physics Reports, 229  (1993) 145.

\noindent
{\bf [20]} The mass fraction of $^4$He, $Y_p$, must satisfy $Y_p$
$^<_{\sim} 0.24$. It is believed that the primordial $^7$Li
abundance is the Population II result $^7$Li/H$\approx 10^{-10}$,
but it is certainly smaller than the Population I value of $\sim
10^{-9}$. See discussions in A. M. Boesgaard and G. Steigman,
Ann. Rev. Astron. Astrophys. 23 (1985) 319; Walker et al., Astrophys. J.,
376 (1991) 51; and ref. [18].

\vfill\eject

\centerline{\bf Figure Captions}
\vskip 0.15in
{\bf Figure 1:} The co-moving proton diffusion length, $d_{100}$,
at temperature $T=500$ keV
as a function of fluctuation amplitude
$(1+\Delta )$. The calculation assumes $\Omega_b h^2=0.0125$.

\vskip 0.15in
{\bf Figure 2:} The $^4$He mass fraction (upper panel) and $^7$Li
number fraction relative to hydrogen from inhomogeneous
nucleosynthesis calculations are shown as functions of the
co-moving fluctuation separation $l_{100}^s$. These calculations employ an
initial density contrast $\Lambda =1.25\times 10^6$, Gaussian width
$a_{100}=l_{100}^s/20$, and $\Omega_bh^2=0.0125$. The dotted lines
indicate the minimum scale above which fluctuations survive baryon
diffusion, BDL, and the electroweak horizon scale, EWH.

\vskip 0.15in
{\bf Figure 3:} Constraints from inhomogeneous big bang
nucleosynthesis on models with a high-density region
volume filling fraction $f_V$ and density contrast $\Lambda
=n_b^H/n_b^L$. Any models with parameters falling to the right
of the shaded line will
overproduce $^4$He relative to observational limits.
For this calculation we
assumed spherical square-wave fluctuations with fluctuation mean separation
$l_{100}^s=0.5$ cm and $\Omega_b h^2=0.0125$.

\end